\documentclass[a4paper]{jpconf}

\usepackage{float}

\usepackage{graphicx}

\usepackage{amsmath, amsthm, amssymb, amsfonts}

\usepackage{bm}% bold math
\usepackage[percent]{overpic}
\usepackage{color}

\def\eg{\eta_{\rm gap}}

\def\pt{p_{\rm T}}

\def 	\bcor{b_{\rm corr}}
\def 	\C2{C_{\rm 2}}

\def\sNN{\mbox{$\sqrt{s_{_{\rm NN}}}$}}

\def\vf{\varphi}

\begin{document}
\title{Long-range correlations in ALICE at the LHC}

\author{I G Altsybeev$^1$  (for the ALICE collaboration)}
\address{$^1$ 
%Saint-Petersburg State University, 7/9 Universitetskaya Emb., 199034, Saint Petersburg, Russia
Saint-Petersburg State University, 7/9 Universitetskaya nab., St. Petersburg, 199034 Russia
}

\ead{i.altsybeev@spbu.ru}

\begin{abstract}
Long-range correlations between particles separated by a pseudorapidity gap are a powerful tool to explore the initial stages and evolution of the medium created in hadron-hadron collisions.  An overview of the long-range correlations measured by the ALICE detector in  pp, p-Pb and Pb-Pb will be presented.  This includes analyses of forward-backward, two- and multi-particle correlations with the use of the central barrel and forward detectors.  
%Comparisons to existing models will be also discussed.
\end{abstract}

\section{Introduction}

Long-range correlations (LRC)
are usually considered between particles separated by pseudorapidity gap,
that is typically taken to be $|\Delta\eta| \gtrsim 1.0$
in order to suppress contribution from resonances and (mini)jets.
%What are the sources of LRC?
LRC can be created only at the early stages of the collision~\cite{dumitru} 
%geometry
and arise in  Color Glass Condensate \cite{cgc} 
and string fusion \cite{string_fusion} models.
%, 
%interactions between strings \cite{dumitru},
%and 
On later stages of the system evolution
they can be modified by medium and final state interactions
(hydrodynamic expansion,
energy loss in medium,
conservation laws).

How do we extract information about long-range correlations?
Different analysis methods are being used, which 
have different sensitivity to various aspects of physical origin of LRC.
%allow to probe di
Two-particle correlations with large separating $\eta$ gaps between pairs of particles allows one
to get rid of short-range correlations and 
obtain the correlation pattern, %$\vf$ 
azimuthal profile of which can be decomposed
into Fourier series.
%Fourier decomposition of azimuthal profile at large $\Delta\eta$:
%elliptic and triangular flow
Another technique, which is sensitive to collective phenomena, uses
multiparticle correlations, for example, elliptic flow can be measured 
by calculating 
 second-order Fourier coefficient
$v_2$  taking 4, 6, 8 or even all particles in event 
(usually denoted as $v_2\{4\}, v_2\{6\} , v_2\{8\} , v_2\{\rm LYZ, \infty\}$). 
Even with 4 particles, %and more particles  allow to suppresses non-flow contribution enough,
non-flow contribution is already suppressed enough,
allowing one to measure collective effects during evolution of the system created in a hadronic collision.
Forward-backward correlation analysis is another method sensitive to even-by-event fluctuations
of number and properties of particle-emitting sources elongated in rapidity.

LRC are measured in Pb--Pb, p--Pb and pp collisions
in all major experiments at the LHC.
This article collects 
some experimental highlights on LRC from ALICE.
%some highlights from the ALICE collaboration
%on LRC studies in
%are under investigation at LHC.

%\section{How do we extract information about long-range correlations?}

\section{ALICE detector}
%\section{Introduction}

In ALICE experimental setup, % consists of many detecting systems.
charged primary particles are reconstructed with the central barrel detectors combining information from the Inner Tracking System (ITS) and the Time Projection Chamber (TPC). Both detectors are located inside 
the ALICE solenoid with a field of 0.5 T
%the 0.5~T solenoidal field
and have full azimuthal coverage for track reconstruction 
within a pseudo-rapidity window of $|\eta| < 0.9$.
%The ALICE TPC is the main tracking detector of the central rapidity region. The TPC, together with the ITS, provides charged particle momentum measurement, particle identification and vertex determination with good momentum and $dE/dx$ resolution.
Charged particle identification (PID) over a broad momentum range 
is done using information from the TPC and the Time of Flight (TOF) detectors.
The TPC provides a simultaneous measurement of the momentum of a particle and its specific ionisation energy loss (d$E/$d$x$) in the gas.
%for the hadron species at pT < 0.7 GeV/c and the possibility to identify particles on a statistical basis in the relativistic rise region of dE/dx (i.e. 2 < pT < 20 GeV/c)
The TOF detector surrounds the TPC and provides  separation between 
$\pi$--K and K--p up to $\pt = 2.5$ GeV/$c$ and $\pt = 4$ GeV/$c$, respectively,
 %[68]
by measuring the arrival time of particles. % \cite{alice_performance}. % with a resolution of about 80 ps.
VZERO detectors,
two forward scintillator arrays
with coverage $-3.7<\eta<-1.7$ and $2.8<\eta<5.1$,
are used in the trigger logic and for the centrality and reference flow particle determination
 \cite{alice_performance}.
%centrality + triggering

%Tracking: ITS+TPC ($\pt > 0.2$ GeV/$c$)
%Particle identification: ITS+TPC+TOF
%Centrality estimators: VZERO, ZDC

\section{Two-particle correlations with pseudorapidity gap}

\begin{figure}[H]
\centering
%\subfigure[a][]
%{
%\begin{overpic}[width=0.48\textwidth,%trim={0.1cm 1.0cm 1.4cm 0.2cm},clip
%]{plots/Harmonic_decomposition_PbPb_2012-Jun-06-fig02a.eps} 
 %\end{overpic}
%}
%\hspace{-0.4cm}
%\subfigure[a][]
%{
\includegraphics*[width=0.54\textwidth]
%, trim={0.1cm 0.0cm 1cm 0.1cm},clip]
{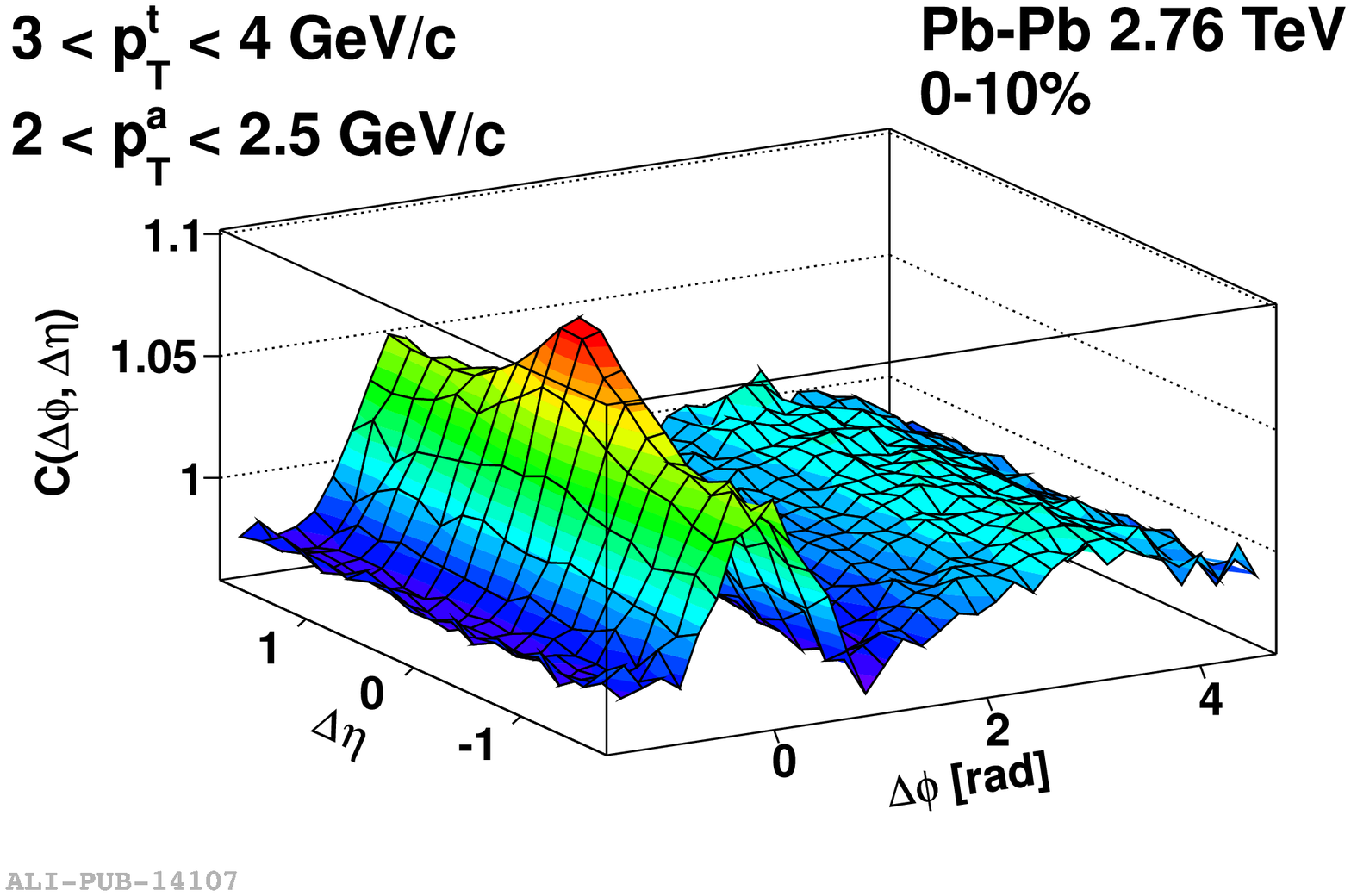} 
%\end{overpic}
%} 
\hfill %\hspace{-0.4cm}
\includegraphics*[width=0.42\textwidth, %trim={0.1cm 0.0cm 0cm 0.1cm},clip
]{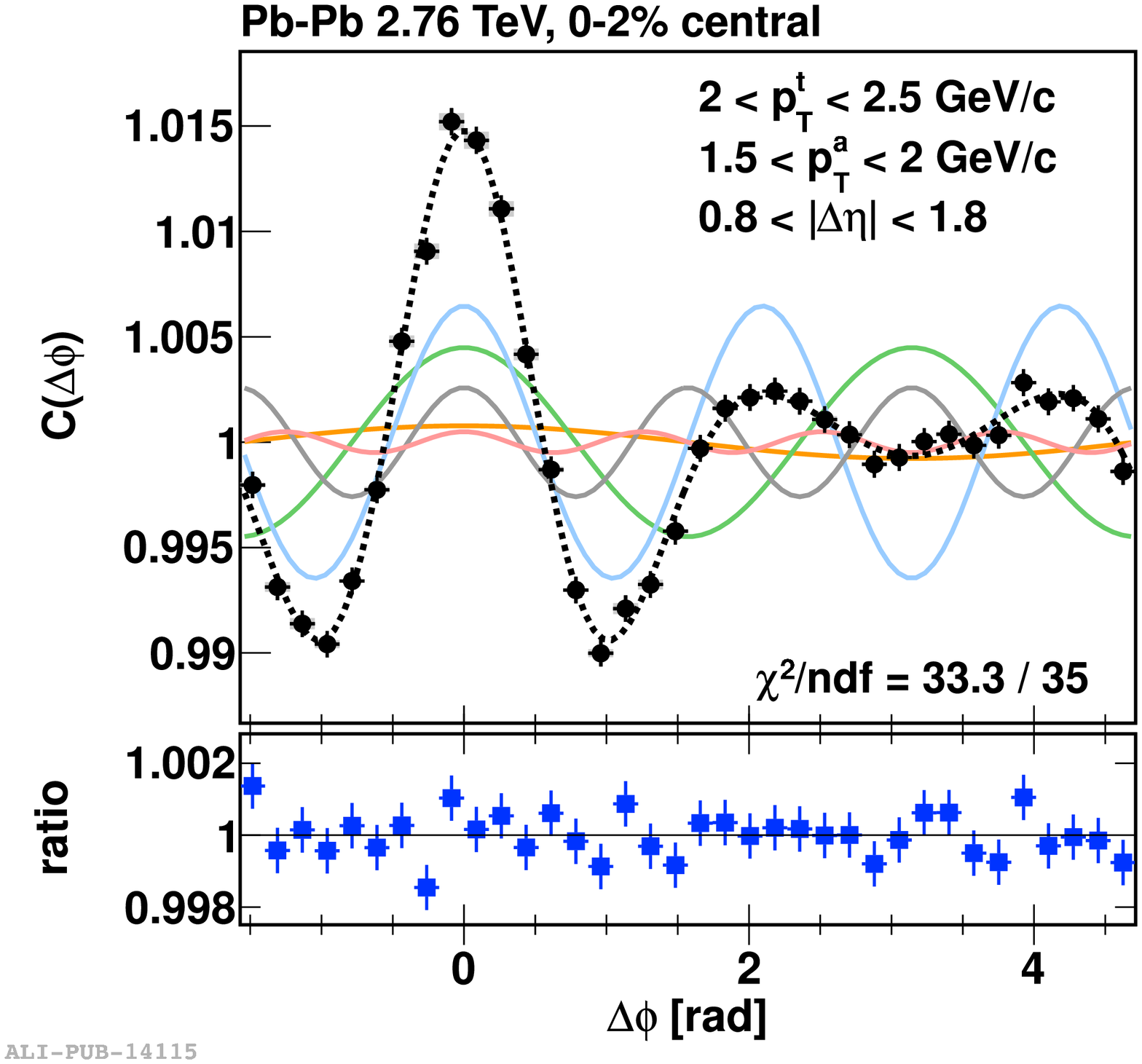} 
\caption{
Left: two-particle correlation function $C(\Delta\eta, \Delta\vf)$
for central 0-10\% Pb-Pb collisions at 2--4 $\pt$ range.
Right: $C(\Delta\vf)$ for particle pairs at $|\eta|>0.8$ in most central 0-2\% events.
The Fourier harmonics for $V_1$ to $V_5$ are superimposed in color \cite{harmonic_decomposition}.
}
\label{2PC_PbPb}
\end{figure}

The two-particle correlation function $C(\Delta\eta, \Delta\vf)$ 
measured in central 0-10\% Pb-Pb collisions, is shown in figure \ref{2PC_PbPb} (a),
demonstrating such structures as  ``near-side" peak from jets and resonances, %HBT),
``near-side ridge" along $\Delta\eta$ at $\Delta\vf\approx 0$ and broad away-side structure elongated in $\Delta\eta$ \cite{harmonic_decomposition}. 
%pseudorapidity-separated ($|\eta| > 0.8$)
The {\it measured} anisotropy of particle production at $n$-th harmonic order %,
%which
 is given by coefficient $V_{n\Delta}$, %(\pt^{trig},\pt^{assoc})$,
which can be obtained from triggered, pseudorapidity-separated ($|\eta| > 0.8$)
pair azimuthal correlations  using  expression
$\frac{dN^{\rm pairs}}{d\Delta\vf}=1+2\sum_{n=1}^{\infty}V_{n\Delta}(\pt^t,\pt^a)\cos(n\Delta\vf)$, 
where $\pt^{t}$ and $\pt^{a}$ are transverse momenta of trigger and associated particles.

In very central events 0-2\% (figure \ref{2PC_PbPb}, b), the away side exhibits a concave,
doubly-peaked feature ($V_{3\Delta}>V_{2\Delta}$), that is usually attributed to fluctuations in
initial state geometry which can generate higher-order flow components.
%“Double hump” structure appears because of larger $v_3$ relative to $v_2$ in most central collisions.
It was checked whether a set of single-valued points
$v_n(\pt)$  can be identified that describe the measured long-range
anisotropy via the relation $v_n(\pt^t)v_n(\pt^a)=V_{n\Delta}(\pt^t,\pt^a)$. If
so, $V_n\Delta$ is said to factorize into single-particle Fourier coefficients.
These pair anisotropies are found to approximately factorize into
single-particle harmonic coefficients for $\pt^a < 4$ GeV/$c$.
This factorization is consistent with
%expectations from 
the picture of collective response of the medium to anisotropic initial conditions.

\section{ $v_2$ for identified particles in different $\pt$ ranges }

\begin{figure}[H]
 %   \begin{center}
 \centering
%\subfigure[a][]
%{
\begin{overpic}
[width=0.48\textwidth,trim={0.1cm 1.0cm 1.4cm 0.2cm},clip]{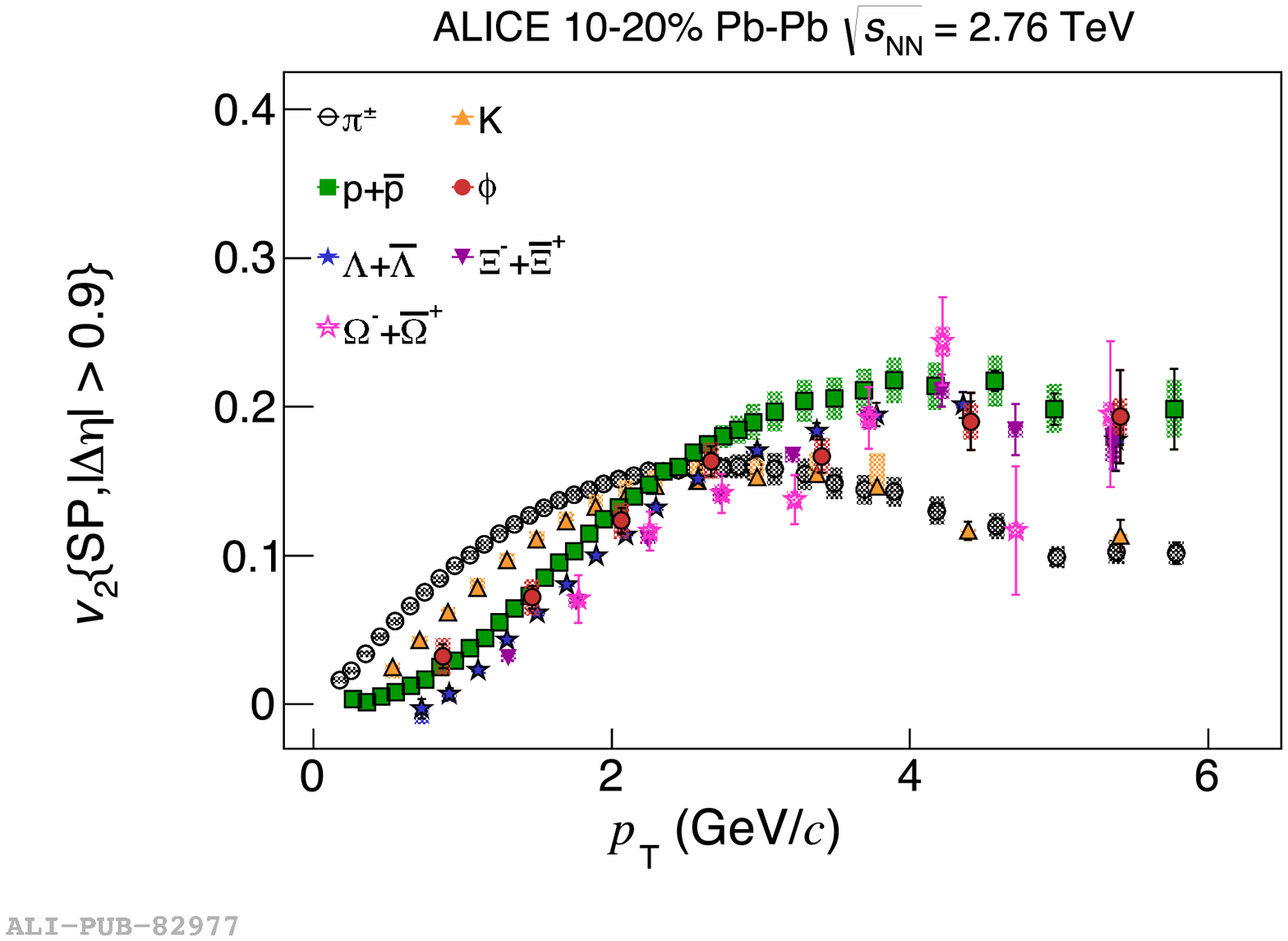} 
  %   \put(25,57){  10\% }
 %    \put(25,45){  5\% }
     \put(1,1){\tiny  ALI-PUB-52116 }
  %    \put(2,102){  \color{blue} smearing of $\pt$ by 1\% }
\end{overpic}
%}
 \hfill %\hspace{-0.4cm}
%\subfigure[a][]
%{
\includegraphics*[width=0.48\textwidth]{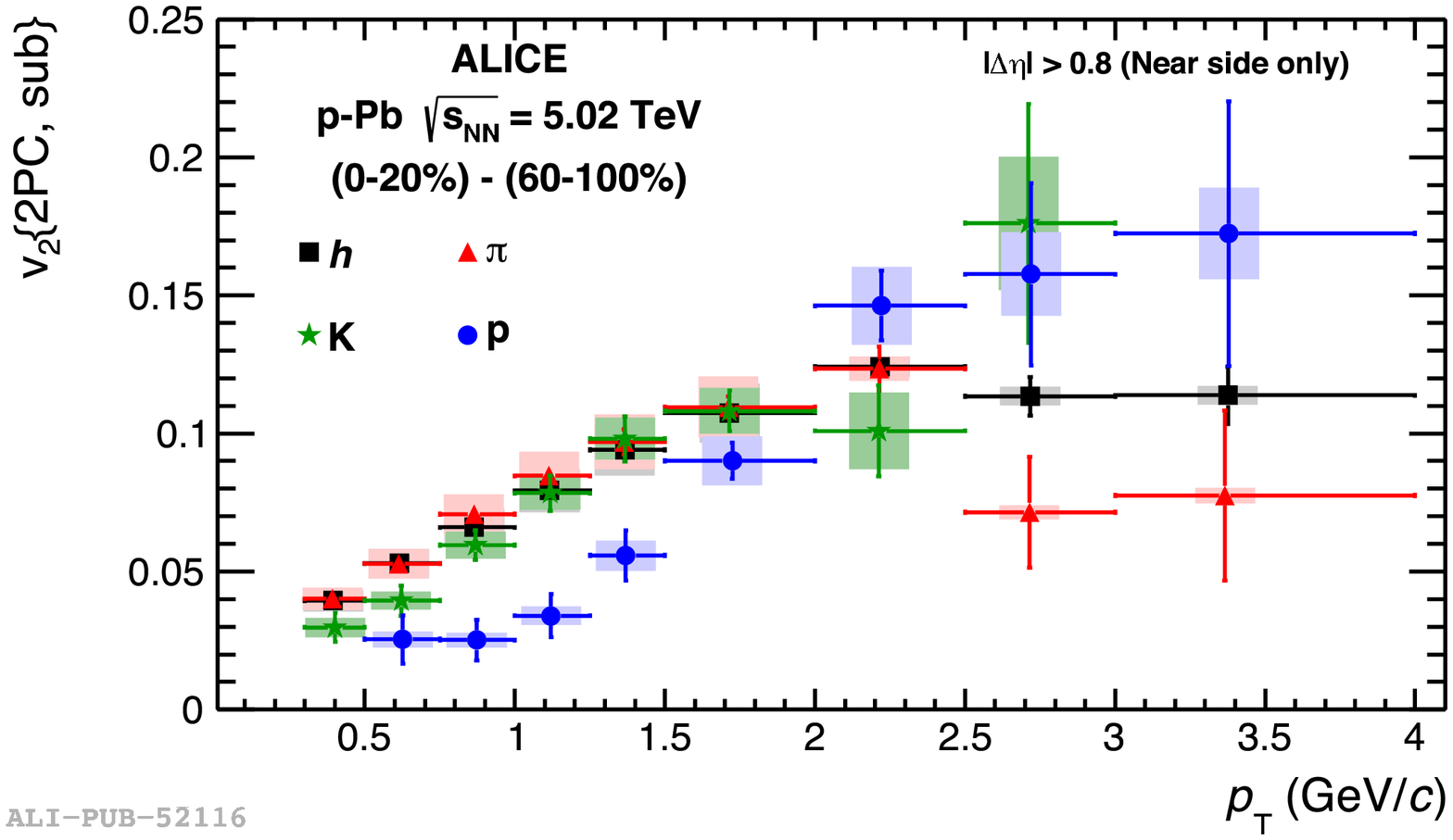} 
%} 
  %  \end{center}
\caption{
The $\pt$-differential $v_2$ for different particle species 
in centrality class 10-20\% of Pb-Pb collisions at $\sNN=2.76$ TeV 
\cite{v2_PID_PbPb} (left)
and in 0-20\% class with subtracted 60-100\% class
of p-Pb collisions at $\sNN=5.02$ TeV \cite{v2_PID_pPb} (right).
}
\label{fig:v2}
\end{figure}

Elliptic flow ($v_2$) in Pb-Pb collisions was measured
with the Scalar Product method, which is a two-particle correlation technique, 
using a pseudo-rapidity gap of $|\eta|>0.9$
between  reference particles in VZERO scintillators
and the identified hadrons of interest within TPC~\cite{v2_PID_PbPb}. % under study and the reference particles. 
Dependence of $v_2$ on $\pt$ in centrality class 10-20\% is shown 
in figure \ref{fig:v2} (a) for $\pi$, K, p, %$\rm K^{\pm}$, p, $\rm K^0_S$, 
$\Lambda$, $\phi$, 
$\Xi$ and $\Omega$. %, demonstrating 
Shift of the $v_2(\pt)$ dependences towards higher $\pt$
is observed for heavier particles due to interplay between elliptic and radial flow
(``mass ordering") up to 3 GeV/$c$. % observed for many species,
Mass ordering is stronger in most central collisions  because of  stronger radial flow.
%Crossing between proton and pion $v_2$ is around $\pt\approx 3$ GeV/$c$
%Baryon/meson splitting persists out to high pT.
%For higher values
%of $\pt$ 
Particles with
$\pt>3$~GeV/$c$ %) %, particles 
tend to group according to their type, i.e. mesons and baryons (except for $\phi$ meson).
In this range of $\pt$,
quark coalescence was suggested to be the dominant hadronization mechanism,
 however, ALICE data exhibit deviations from the number of constituent quark (NCQ) scaling at the level of $\pm 20\%$.

Mass ordering was  observed also for $\pi$, %$\rm K^{\pm}$, 
K and p in p-Pb collisions,
when $v_2$ was extracted from two-particle correlations in high-multiplicity events
after subtraction of correlations in low multiplicity events (figure \ref{fig:v2}, b) \cite{v2_PID_pPb}.
Qualitatively similar picture in p-Pb as in Pb-Pb collisions
suggests common origin of long-range correlations %that flow exists not only in A-A but also in smaller systems,
in A-A and in smaller systems.
This question currently is under intensive debates and investigation.

\section{ “Forward-central” two-particle correlations in p-Pb at $\sNN=5.02$ TeV}

Further studies of long-range correlation structures were performed in p-Pb collisions using
inclusive muons from muon spectrometer as trigger particles
(with $0.5 < \pt < 4$ GeV/$c$,  $-4 < \eta < -2.5$)
and associated charged particles in ITS+TPC ($|\eta| < 1$, $0.5 < \pt < 4$ GeV/$c$)~\cite{v2_muon_track_pPb}.
Reconstructed muons mainly originate from weak decays of $\pi$, K and mesons from heavy flavor (HF) decays.

\begin{figure}[t]
\centering
\includegraphics*[width=0.38\textwidth,trim={0.0cm 0.0cm 0cm 0cm},clip]{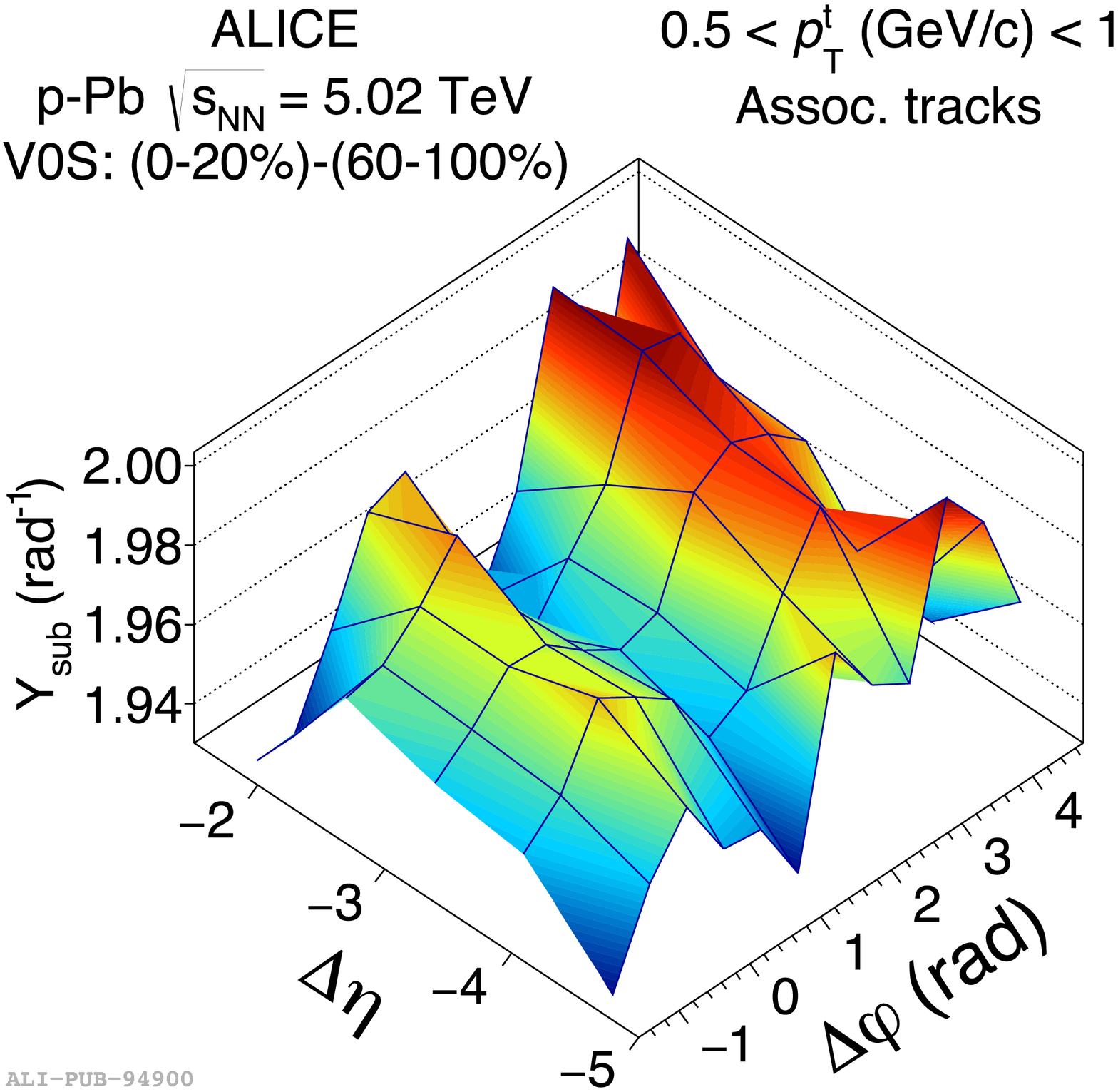} 
\hspace{0.4cm}
\includegraphics*[width=0.43\textwidth,trim={0cm 0cm 0cm 0.0cm},clip]{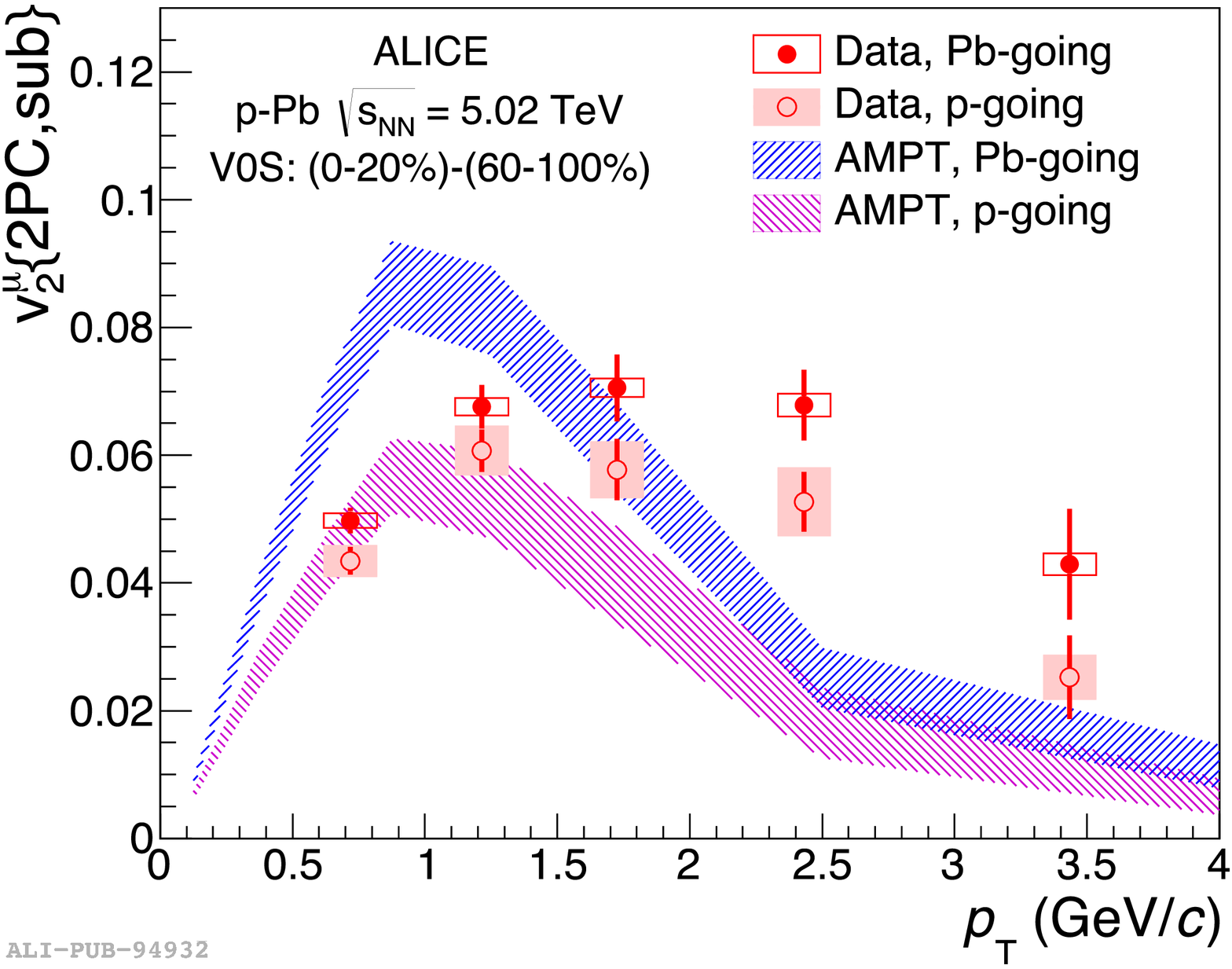} 
\caption{
Left: associated yield per trigger particle 
in 0-20\% event classes after subtraction of 60-100\% class
as a function 
$\Delta\eta$ and $\Delta\vf$ for muon-track correlations in p-Pb.
Right: the $v_2^\mu\rm \{2PC, sub\}$ coefficients from muon-tracklet correlations in p-going and Pb-going directions~\cite{v2_muon_track_pPb},
the data are compared to model calculations from AMPT.
}
\label{fig:muon_track}
\end{figure}

Figure \ref{fig:muon_track} (a) shows yield of associated particles in $(\Delta\eta, \Delta\vf)$
coordinates in 0-20\% event classes after subtraction of 60-100\% class.
%associated yield per trigger particle in 0-20% event classes after subtraction of 60-100% class
%Results:
It can be seen that near-side ridge extends up to $|\Delta\eta| \approx 5$ and 
%$|\eta|\approx 4$,
it decreases from 􏱖$\eta = -1.5$ to $\eta = -5.0$.
Analysis was done at  both beam directions p-Pb and Pb-p, % were analyzed.
and,
assuming factorization of
the Fourier coefficients at central and forward rapidity,
 extracted $v_2$ coefficients %were extracted
%and, 
were found to be larger by $16 \pm 6$ 
when muons are measured in Pb-going direction,
rather than p-going (figure \ref{fig:muon_track}, b). 
Calculations using AMPT event generator showed qualitatively similar behavior at low $\pt$.

%Pb-going coefficients larger by $16 \pm 6$ than p-going,
%comparison with AMPT: similar behavior at low $\pt$

\if 0
trigger particles = inclusive muons:  $0.5 < \pt < 4$ GeV/$c$ 
at  $-4 < \eta < -2.5$,
associated particles = charged particles in ITS+TPC: 
$|\eta| < 1$ at $0.5 < \pt < 4$ GeV/$c$
(or “tracklets” in SPD: $\pt \ge$ approx 50 MeV/$c$)
\fi

%Associated yield per trigger muon (...).

%(similar measurements p-Pb and Pb-p beam directions)
%Assume factorization: (...)

%\vspace*{-0.3cm}
\section{ Forward-backward multiplicity correlations in pp collisions }

\begin{figure}[H]
\centering
\includegraphics*[width=0.82\textwidth]{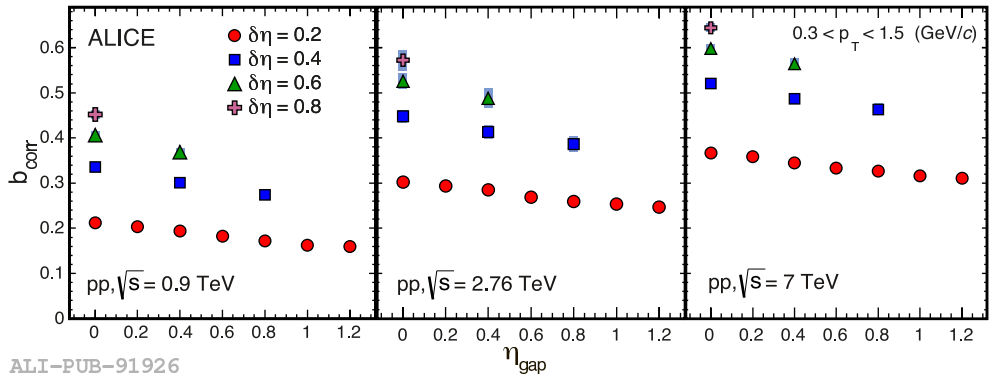} 
\caption{
Forward-backward correlation strength $\bcor$ as function of $\eg$  
and for different windows widths  $\delta\eta$=0.2, 0.4, 0.6 and 0.8 in pp collisions  at $\sqrt{s}=0.9$, 2.76 and 7~TeV \cite{FB_pp}.
}
\label{fig:fullPhiAllWindows}      
\end{figure}

The forward-backward (FB) multiplicity correlations were previously studied
%were studied experimentally 
in a large number of colliding systems.
%Analysis in 
These correlations are characterized by the correlation strength $\bcor$.
ALICE  performed analysis of FB multiplicity correlations in pp collisions in a soft range of $\pt$ (0.3--1.5 GeV/$c$)
at $\sqrt{s}=0.9$, 2.76 and 7~TeV \cite{FB_pp}. % in pp collisions.
Figure \ref{fig:fullPhiAllWindows}  demonstrates behaviour of $\bcor$ 
as a function of separating $\eta$ gap between forward and backward windows.
The correlation strength drops with $\eta$ gap since short-range contribution reduces,
also $\bcor$ values increase with collision energy.

\begin{figure}[t]
\centering
\includegraphics*[width=0.43\textwidth]{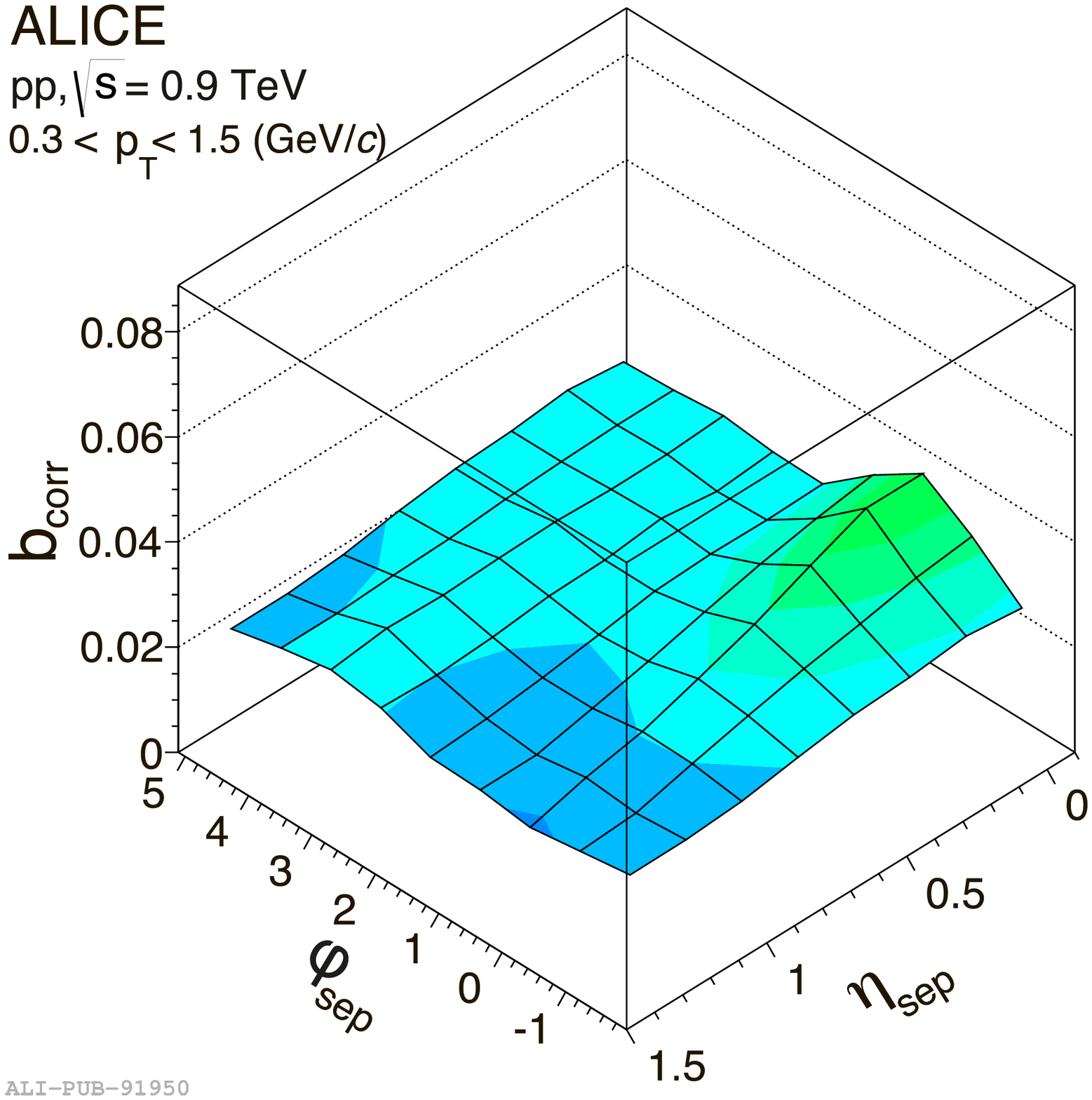}
\hspace{0.4cm}
\includegraphics*[width=0.43\textwidth]{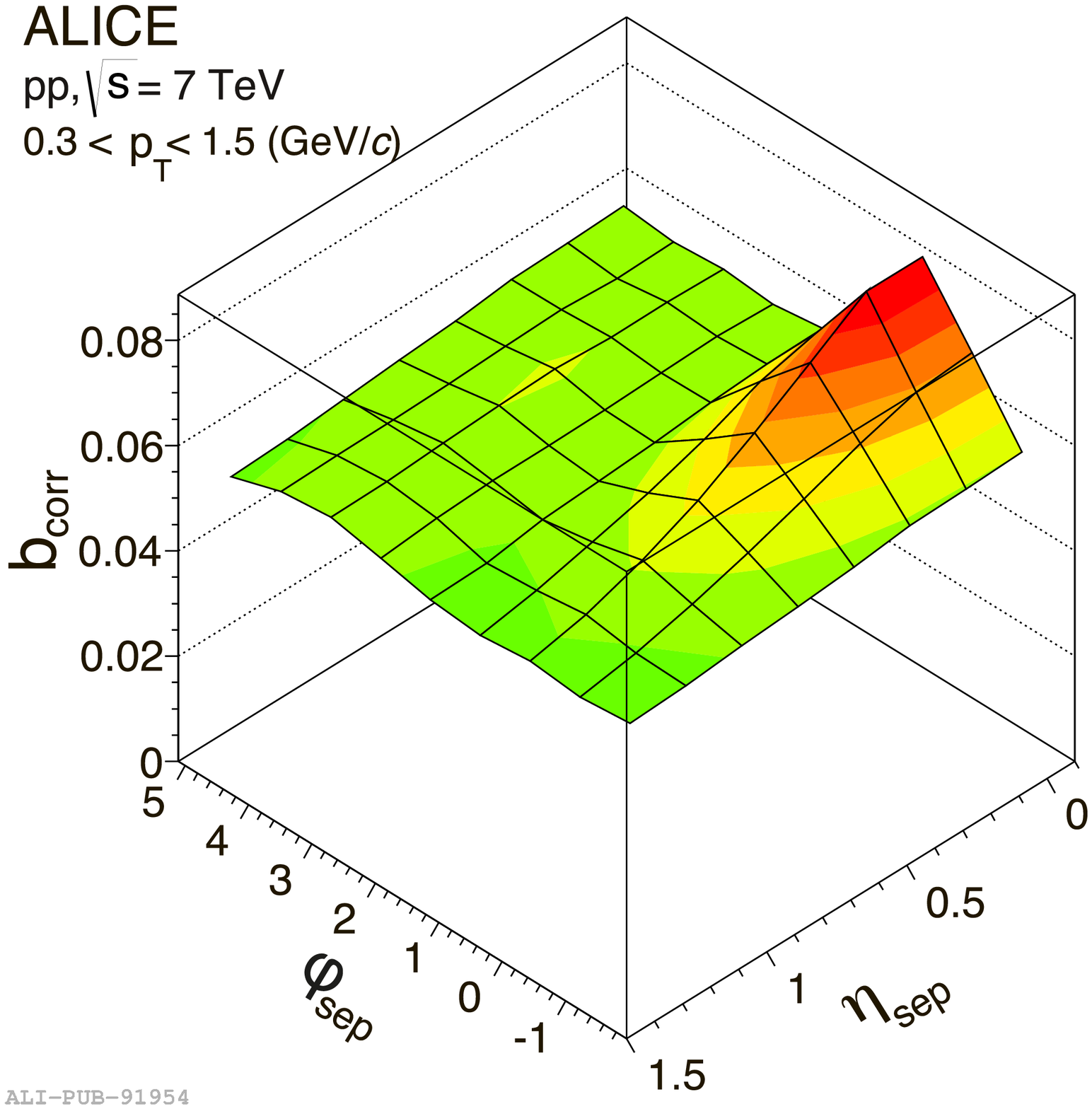}
\caption{
2D representation of $\bcor$ values in forward-backward $\eta$-$\vf$
window pairs with widths $\delta\eta$=0.2 and $\delta\varphi$=$\pi$/4
%of  $\bcor$ 
at    $\sqrt{s}=0.9$~TeV (left)
and 
 at  $\sqrt{s}=7$~TeV (right)
%for separated $\eta$-$\varphi$  
\cite{FB_pp}.
%To improve visibility, the point ($\esep$,$\fsep$)=(0,0) and thus $\bcor$=1 is limited 
%to the level of the maximum value in adjacent bins.
}
\label{fig:pp_3DbothEnergies_8by8}
\end{figure}

In order to get more insight into particle production mechanisms,
conventional analysis of the FB multiplicity correlations between two $\eta$-intervals was extended  
into azimuthal dimension. %, providing additional insight into particle production mechanisms.
Figure \ref{fig:pp_3DbothEnergies_8by8} for two energies $\sqrt{s}=0.9$ and  7~TeV
shows that correlation pattern in $\eta$-$\vf$ space can be split into short-range peak and 
{\it non-zero plateau} which spreads over all studied $\eta$-$\vf$ separations of FB windows 
(underlying long-range correlation).
%short-range and long-range contributions are distinguishable,
This non-zero plateau is found %and 
to increase with the collision energy.
In a simple model of independent particle emitters \cite{FB_VV} this rise  %can be
%is interpreted as increase of
is attributed to increase of fluctuations in number of independent sources % independent particle emitters 
(strings).

%\vspace*{-1.2cm}
\section{ Final remarks }
Long-range correlations provide insights into initial stages and details of the evolution of hadronic collisions.
The correlations are being studied at ALICE in pp, p-Pb, Pb-Pb collisions by several analysis methods.
%Distant regions of $\eta$ have common velocity field
%which is supported by 
Two-particle correlations with separating $\eta$ gap reveal azimuthal correlation
structures, like double-hump shape in central Pb-Pb collisions.
Clear mass ordering at $\pt \lesssim 3$~GeV/$c$ in p-Pb, Pb-Pb events 
%manifests
%that distant regions of $\eta$  range are connected via a common velocity field.
%, p-Pb collisions: 
%Similar behaviour in Pb-Pb and p-Pb is seen also for $v_3$.
%These observations are 
is consistent with the picture of hydrodynamic expansion of the medium in A-A collisions,
and also states the question about origin of the  flow in small systems. % and its origin? 
%p-Pb collisions: similar features for v2 and v3 as in Pb-Pb 
	%-- flow in small systems? 
Analysis of muon-hadron correlations in p-Pb collisions shows that long-range double-ridge 
persists up to $\Delta\eta\approx 5$.
Forward-backward correlations in pp  provide access to fluctuations in number 
and properties of particle-emitting sources.

With the limited central barrel acceptance for tracking, 
ALICE can measure long-range correlations with other, non-tracking detecting systems.
For example, % recently analysis of 
$\eta$-dependence of the anisotropic flow in Pb-Pb collisions
was measured recently with two- and four-particle correlations 
using forward detectors in a range $-3.5<\eta<5$
\cite{dep_vn_on_eta}.
%Another recent analysis of Pb-Pb collisions at $\sNN=5.02$ TeV shows that  increase of anisotropic integrated flow from 2.76 to 5.02 TeV is very moderate
%\cite{v2_502}.
After upgrade in 2019--2020,  ALICE will have a new Inner Tracking System with larger $\eta$ coverage
 and also a new Muon Forward Tracker (MFT).
Combined ITS+MFT acceptance %$\eta$-acceptance 
$-3.6<\eta<1.2$
will significantly extend possibilities to study LRC
of charged particles. % with $\eta$-separation up to 5 units.
%With upgraded ALICE, long-range correlations can be studied in much broader rapidity range.

\if 0
New ITS:
 inner layers: $|\eta|<2.0$ 
 outer layers: $|\eta|<1.3$
If use >=4 ITS layers and >=4 MFT disks:
$\eta$-separation up to 5 units in $|z| < 10$ cm
with a minimum a $\eta$ gap of 1 unit

\fi

%\vspace*{-0.2cm}

%\vspace*{-0.5cm}
\section*{Acknowledgements}
This work is supported by the Saint-Petersburg State University research grant 11.38.242.2015. 

%\vspace*{-0.2cm}


\section*{References}
\begin{thebibliography}{9}

\bibitem{dumitru}
Dumitru A {\it et al.} 2008
{\it Nucl. Phys.}  A {\bf 810} 91–108


\bibitem{cgc}
McLerran L 2002
{\it Nucl. Phys.} A {\bf 699} 73 


\bibitem{string_fusion}
Braun M A  {\it et al.} 1992
{\it Phys. Lett.} B {\bf 287} 154-158
%Phys. B390 (1993) 542, 549



\bibitem{alice_performance}
Abelev B B {\it et al.} 2014, 
{\it Int. J. Mod. Phys.} A {\bf 29} 1430044

\bibitem{harmonic_decomposition}
%ALICE Collaboration,
% Harmonic decomposition of two-particle angular correlations in Pb-Pb collisions at $\sNN = 2.76$ TeV,
Aamodt K {\it et al.} 2012
{\it Physics Letters} B {\bf 708}  249-264

\bibitem{v2_PID_PbPb}
%ALICE Collaboration,
%Elliptic flow of identified hadrons in Pb-Pb collisions at $\sNN=2.76$ TeV,
Abelev B B {\it et al.} 2015
{\it JHEP} {\bf 06} 190


\bibitem{v2_PID_pPb}
%ALICE Collaboration,
%Long-range angular correlations of $\pi$, K and p in p-Pb collisions at $\sNN=5.02$ TeV,
Abelev B B {\it et al.} 2013
{\it Phys. Lett.} B {\bf 726} 164-177

\bibitem{v2_muon_track_pPb}
%ALICE Collaboration,
%Forward-central two-particle correlations in p-Pb collisions at $\sNN=5.02$ TeV,
Adam J {\it et al.} 2016
{\it Phys. Lett.} B {\bf 753}  126-139


\bibitem{FB_pp}
%ALICE Collaboration,
%Forward-backward multiplicity correlations in pp collisions at $\sqrt{s}$ = 0.9, 2.76 and 7 TeV,
Abelev B B {\it et al.} 2015
{\it JHEP} {\bf 1505}  097 

\bibitem{FB_VV}
Vechernin V V 2015
{\it Nucl.Phys.} A {\bf 939} 21 %, arXiv:1210.7588.


\bibitem{dep_vn_on_eta}
%Pseudorapidity dependence of the anisotropic flow of charged particles in Pb-Pb collisions at sqrt{sNN}=2.76 TeV
Adam J {\it et al.} 2016
arXiv:1605.02035


%\bibitem{v2_502}
%ALICE Collaboration, 2016 PRL 116, 132302.



\end{thebibliography}
\end{document}